\documentclass[times,twocolumn,final,authoryear]{elsarticle}

\usepackage{jasr}
\usepackage{framed,multirow}

\usepackage{amssymb}
\usepackage{latexsym}

\usepackage{url}
\usepackage{xcolor}
\definecolor{newcolor}{rgb}{.8,.349,.1}

\usepackage[citebordercolor=white]{hyperref}

\journal{Advances in Space Research}

\begin{document}

\verso{C. Mac Cormack \textit{et. al}}

\begin{frontmatter}

\title{Enthalpy-based modeling of \\
tomographically reconstructed quiet-Sun coronal loops}

\author[1,2]{C. \snm{Mac Cormack}\corref{cor1}}
\cortext[cor1]{Corresponding author}
\ead{cmaccormack@iafe.uba.ar}
\author[1]{M. \snm{L\'opez Fuentes}}
\author[1]{C. H. \snm{Mandrini}}
\author[1]{D. \snm{Lloveras}}
\author[1,2]{A. M. \snm{V\'asquez}}

\address[1]{Instituto de Astronom\'{\i}a y F\'{\i}sica del Espacio (IAFE, UBA-CONICET), Casilla de Correo 67 Suc. 28, 1428, Ciudad Aut\'onoma de Buenos Aires, Argentina.}
\address[2]{Departamento de Ciencia y Tecnolog\'{\i}a, Universidad de Tres de Febrero, Valentín Gómez 4828, 1678, Caseros, Argentina.}

\received{3 Mar 2022}
\accepted{1 Apr 2022}
\availableonline{7 Apr 2022}

\begin{abstract}

The structure of the solar corona is made of magnetic flux tubes or loops. Due to the lack of contrast with their environment, observing and studying coronal loops in the quiet Sun is extremely difficult. In this work we use a differential emission measure tomographic (DEMT) technique to reconstruct, from a series of EUV images covering an entire solar rotation, the average 3D distribution of the thermal properties of the coronal plasma. By combining the DEMT products with extrapolations of the global coronal magnetic field, we reconstruct coronal loops and obtain the energy input required to keep them at the typical million-degree temperatures of the corona. We statistically study a large number of reconstructed loops for Carrington rotation (CR) 2082 obtaining a series of typical average loops of different lengths. We look for relations between the thermal properties and the lengths of the constructed typical loops and find similar results to those found in a previous work \citep{maccormack_2020}. We also analyze the typical loop properties by comparing them with the zero-dimensional (0D) hydrodynamic model Enthalpy-Based Thermal Evolution of Loops \citep[EBTEL, ][]{klimchuk_2008}. We explore two heating scenarios. In the first one, we apply a constant heating rate assuming that typical loops are in quasi-static equilibrium. In the second scenario we heat the plasma in the loops using short impulsive events. We find that the reconstructed typical loops are overdense with respect to quasi-static equilibrium solutions of the hydrodynamic model. Impulsive heating, on the other hand, reproduces better the observed densities and temperatures for the shorter and approximately semicircular loops. The thermal properties of longer loops cannot be correctly reproduced with the EBTEL model. We suggest that to properly assess the physical characteristics of the analyzed loops in future works, it would be necessary to use a more sophisticated 1D model, with which to study the loop temperature and density profiles and test localized heating at different locations along the loops.

\end{abstract}


\end{frontmatter}


\section{Introduction}
\label{sec1}

Due to its low plasma $\beta$ (the ratio between the plasma pressure to the magnetic pressure), the coronal magnetic field has a dominant role in active regions, restricting transport phenomena along magnetic-flux tubes or loops. Cross-field energy transport is almost completely inhibited as it becomes evident in active region observations in X-ray and extreme-ultraviolet (EUV) wavelengths, where individual loops are easily identified and seen to evolve in an approximately independent manner \citep{reale_2014}. During solar activity minima and outside active regions the solar corona is more homogeneous, making the direct identification and study of individual magnetic structures a more difficult task. Some of the properties of the quiescent coronal plasma can, however, be determined using methods based on the reconstruction of the differential emission measure (DEM) \citep{morgan_2017}.

As we did in previous works \citep{nuevo_2013,lloveras_2017,maccormack_2017}, here we use a DEM tomographic (DEMT) technique to obtain a global description of the solar corona. The technique provides the thermal properties of the solar corona using as input series of EUV images covering a complete solar rotation. By combining the DEMT products with coronal magnetic field extrapolations, it is possible to obtain the distribution of density and temperature along individual loops. One of the most interesting questions that this method can help to respond is which is the heating mechanism that keeps the solar corona at million-degree temperatures, while the photospheric temperature is two orders of magnitude lower. In a previous work \citep{maccormack_2017}, we developed a DEMT tool that assumes energy balance on each loop and computes the energy input flux at the coronal base necessary to maintain the quiescent corona at the obtained temperatures and densities. 

The assumption of energy balance has been the focus of several works aimed at finding scaling laws to determine whether observed loops are in quasi-static equilibrium or not \citep{rosner_1978,vesecky_1979}. These studies, based on X-ray observations of active region loops, found that loops were consistent with a state of equilibrium and that, in this scenario, the three terms of the balance equation (conductive and radiative losses and the injected energy) should be approximately of the same order. Using relations between these energy terms, it is possible to derive approximate equilibrium scaling laws between temperature and density of individual loops \citep[see e.g., ][]{lopezf_2007}. {blue}{It is worth to add that, as early as 1980, \citet{roberts_1980} analyzed and discussed the limitations of the quasi-static scaling laws found by \citet{rosner_1978}. In particular, \citet{porter_1995} and \citet{tsuneta_1995} found alternate scaling laws derived from soft X-ray observations. Later works also compared scaling laws obtained from active region observations with those predicted by different heating models \citep[see e.g., ][]{fisher_1998, mandrini_2000, petsov_2003, jain_2006}.}

{Active region observations in the EUV indicated that loops observed in this wavelength range are too dense to be explained by quasi-static equilibrium \citep{aschwanden_2001, winebarger_2003}. It has been shown that loop evolutions based on impulsive heating \citep[see e.g., ][]{reale_2000, warren_2002} are able to explain the observed overdensities and flat coronal temperatures. Recently, in \citet{nuevo_2020}, we studied a series of active region loops observed in the EUV and confirmed that their properties were more consistent with impulsive heating, conversely to with quasi-static equilibrium. A thorough review of studies based on loop modeling and observations can be found in \citet{reale_2014}.} 

In \citet{maccormack_2020} we studied scaling laws of loops in the quiescent corona and found relations between thermal properties and loop lengths that are different from those deduced by \citet{rosner_1978}. {In that work, we reconstructed tomographically Carrington Rotation (CR) 2082, which occurred during the minimum between Solar Cycles 23 and 24. We analyzed the relation between the loop-average plasma properties and the length of each reconstructed loop finding the mentioned scaling laws. Here, we revisit the same data set but we build typical loops by averaging the thermal properties of 10 subsets of reconstructed loops with similar lengths (see Section~\ref{RL}). We then obtain 10 typical loops whose plasma properties as a function of position along the loop are known. 
Although in \citet{maccormack_2020} we concluded that the reconstructed loops are not in quasi-static equilibrium, here we explore both, if the typical loops confirm the scaling laws found in the previous work and, more importantly, which is the heating mechanism that best reproduces the loop-average density and temperature of the typical loops, in particular, if impulsive heating provides a better explanation for the obtained densities.} 

In order to determine if typical loops are consistent or not with quasi-static equilibrium, we model the loops using two heating scenarios: quasi-static equilibrium by means of constant heating and heating by impulsive events. To do so, we use the 0D hydrodynamic model Enthalpy-Based Thermal Evolution of Loops  \citep[EBTEL, ][]{klimchuk_2008, cargill_2012} whose main input parameters are the loop length and the heating rate as a function of time.

In Section~\ref{DEMT} we briefly describe the data used in the DEMT procedure and a simple energy balance model used to compute the energy flux required to keep the reconstructed loops in equilibrium. A more detailed description of the DEMT technique and the magnetic field extrapolation is provided in \ref{ADEMT}. In Section~\ref{EBTELcap} we describe the EBTEL model. In Section~\ref{RL} we present the method to obtain the typical loops and the study of the scaling laws that they follow. In Section~\ref{EBTELDEMT} we compare the DEMT results and the EBTEL modeling. We discuss and conclude in Section~\ref{conc}.

\section{Data and loop reconstruction}
\label{DEMT}

{We use differential emission measure tomography (DEMT) to reconstruct the plasma properties of the global corona in the height range of $17 - 174\,\rm{Mm}$. A more detailed description of this technique can be found in \ref{ADEMT}.}

{Since the DEMT is a global technique, short timescales are not well resolved. Moreover, active regions and short term dynamic phenomena are not only very difficult to handle, but they can be also counterproductive for a reliable reconstruction of the mean properties of the global corona \citep[see e.g., ][]{vasquez_2015}. For this reason, the tomographic technique is usually applied to Carrington rotations (CRs) that occur during solar minima. Here, we reconstruct CR 2082, observed between April 6 and May 3, 2009, which does not present active regions or other short-term duration phenomena that could affect the results. For the reconstruction we use EUV images from the Extreme Ultraviolet Imager (EUVI) telescope \citep{howard_2008} on board the Solar Terrestrial Relations Observatory (STEREO), with a cadence of one image every 6 h. We use the coronal bands corresponding to 171, 195, and 284 \AA, that have maximum sensitivity in the temperature range $[0.5,2.5]\,\rm{MK}$ \citep{nuevo_2015}.}

{As explained in \ref{ADEMT} DEMT results can be combined with a magnetic field extrapolation to obtain the plasma properties along magnetic field lines which are loop proxies. To construct the potential-field source-surface (PFSS) model \citep{huang_2012} we use a synoptic magnetogram obtained with the Michelson Doppler Imager (MDI), on board the Solar and Heliospheric Observatory \citep[SOHO, ][]{scherrer_1995}.}

Once we have densities and temperatures along individual magnetic lines or loops, we obtain the radiative and conductive fluxes in each loop using the following equations: 

\begin{eqnarray}
  \phi_r = \left(\frac{B_0 B_L}{B_0+B_L}\right)\ \int_{0}^L \rm{d}s\,\frac{E_r(s)}{B(s)}\, \label{phir} \\
  \nonumber \\
  \phi_c = \frac{B_0F_c(L)-B_LF_c(0)}{B_0+B_L}, \label{phic}
\end{eqnarray}

\noindent
where $B_0$ and $B_L$ represent the magnetic field strength at both coronal bases of a field line, and $F_c$ denotes the thermal conductive flux given by Spitzer's equation $F_c(s) = -\kappa T(s)^{5/2} dT(s)/ds$, with $\kappa = 9.2\times10^{- 7}\,\rm{erg}\,\rm{cm}^{- 1}\,\rm{K}^{- 7/2}$, the Spitzer conductivity. {The temperature gradient $dT(s)/ds$ and the basal temperature are computed from a linear fit of the temperature distribution along each reconstructed loop.}

To obtain the radiative power, $E_r$, we use the LDEM and a radiative loss function, $\Lambda_D(T)$, computed with the CHIANTI database and emission model (Dere et al., 1997):

\begin{equation}
 E_r = \int \rm{d}T\,\rm{LDEM}(T)\,\Lambda_D(T) \approx N^2_e\,\Lambda_D(T),
\end{equation}

We compute the energy input flux at the coronal base ($\approx 17\,\rm{Mm}$) assuming an energy balance between fluxes: $\phi_h = \phi_r + \phi_c$, where $\phi_h$ is the heating influx. 

For a full description of the loop reconstruction technique and the energy flux computation we refer the reader to \citet{maccormack_2017}.

\section{EBTEL Model}
\label{EBTELcap}

To determine if the properties of the studied loops are consistent with equilibrium conditions, we use the 0D hydrodynamic model {Enthalpy-Based Thermal Evolution of Loops} (EBTEL), originally developed by \citet{klimchuk_2008} and later improved by \citet{cargill_2012}. The model is based on the enthalpy balance between the transition region and the corona, to provide a temporal evolution of the mean plasma parameters of the loop in both regions.

The EBTEL model solves the time-dependent equation of energy balance for a half semicircular loop with constant cross-section. The loop is split in two parts: a larger coronal part of length $L$, and a smaller part of length $l~(l<<L)$ corresponding to the transition region (TR). Using the piecewise radiative loss function $\Lambda(T)$ described in \citet{klimchuk_2008} and integrating the energy balance equation in each segment, the model provides a relation between the downward conductive flux from the corona and the energy loss by radiation in the TR:

\begin{equation}
\label{EBTELF}
 \frac{5}{2}P_0v_0 \approx -F_0 - \phi_{r,TR}
\end{equation}

\noindent where {$\frac{5}{2}P_0v_0$} and $F_0$ correspond respectively to the enthalpy and the conductive fluxes at the coronal base and $\phi_{r,TR}$ is the radiative loss flux in the TR portion. 

Depending on the direction of the enthalpy flux between both regions, two alternative scenarios are possible. If the downward conductive flux from the corona is larger than the radiating capacity of the TR $(|F_0| > \phi_{r,tr})$, a positive enthalpy flux results and plasma is evaporated into the corona, increasing the density of the coronal portion of the loop. Otherwise, if the energy transferred by the conductive flux is smaller than the energy radiated by the TR $(|F_0| < \phi_{r,tr})$, the enthalpy flux is negative and plasma condenses from the corona to the TR, decreasing the density in the coronal part of the loop. The $|F_0| = \phi_{r,tr}$ situation corresponds to equilibrium. By combining the balance equations for the TR and the solar corona, EBTEL provides the temporal evolution of the mean thermal parameters along each portion of the loop. The main input parameters of the EBTEL model are the half-length of the loop and the coronal volumetric heating rate function. 

\section{Typical loops}
\label{RL}

In this work we consider only closed loops whose apexes are within the tomographic limits. We split all loops in two legs, focusing the study on the behavior of the thermal properties from the base to the top. This is particularly convenient for the comparison with hydrodynamic models since they usually model legs instead of full loops. To be included in the final set, the legs must meet two conditions: {they must have reconstructed data in at least 5 crossed tomography voxels, and those data must be well distributed {along the entire length of the leg} to have a reasonably spread sample. {For this, we split each leg in three equal parts and we require to have at least one data point on each part.}

Under these conditions we complete a data set of $54704$ legs corresponding to loop lengths in the range $[90,1050]\,$Mm. Our main objective is to determine typical loop leg profiles that represent the mean thermal properties of loops with similar lengths. To do so we split the full set of loops in 10 length bins each containing $\approx5500$ legs. 

{In Figure~\ref{L-dist} we present a frequency histogram of the loop length $L$ for all reconstructed loops. Big dots and error bars denote the median value and the standard deviations of each length bin.}

For each length bin, we construct a typical loop leg profile by taking the median density, temperature, and magnetic field of all the legs in the bin at 10 equidistant average positions along the leg. We also compute the standard deviation of these quantities for each data point as error estimations. This averaging of the loop properties is similar to what is usually done in superposed epoch analysis methods \citep{singh_2006}. {One advantage of using typical loops instead of global averages is that this method provides information, in a statistical fashion, about the dependence of loop properties on the position along the loops for quiescent coronal loops of different lengths. This procedure includes the possibity of future 1D hydrodynamic modeling of typical loops (see also Section~\ref{conc}).}

\begin{figure}[ht!]
\begin{center}
\includegraphics[width=0.48\textwidth]{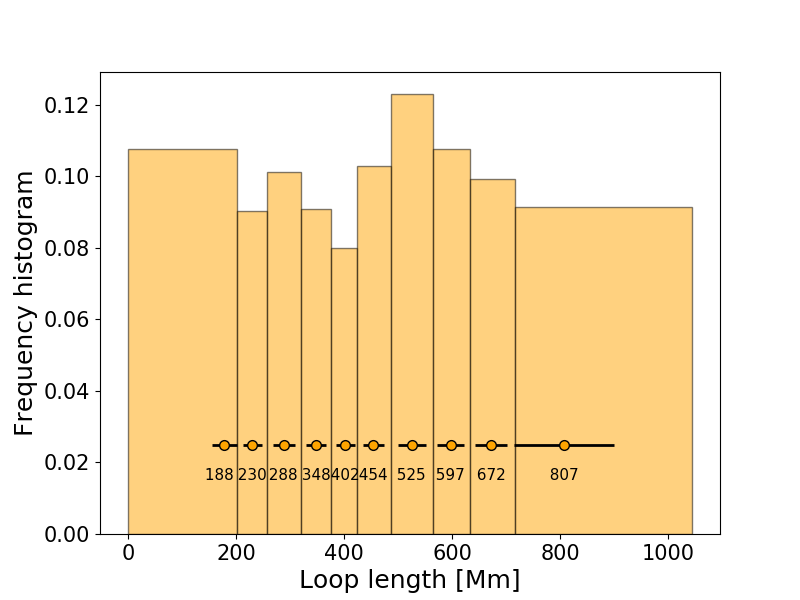}
\caption{Distribution of loop lengths, $L$, for the loops reconstructed with the tomographic technique. The orange circles represent the median value of each length bin and the black bars correspond to the standard deviations taken as a measure of the errors.}
\label{L-dist}
\end{center}
\end{figure}

\begin{figure*}[ht!]
\begin{center}
\includegraphics[width=0.96\textwidth]{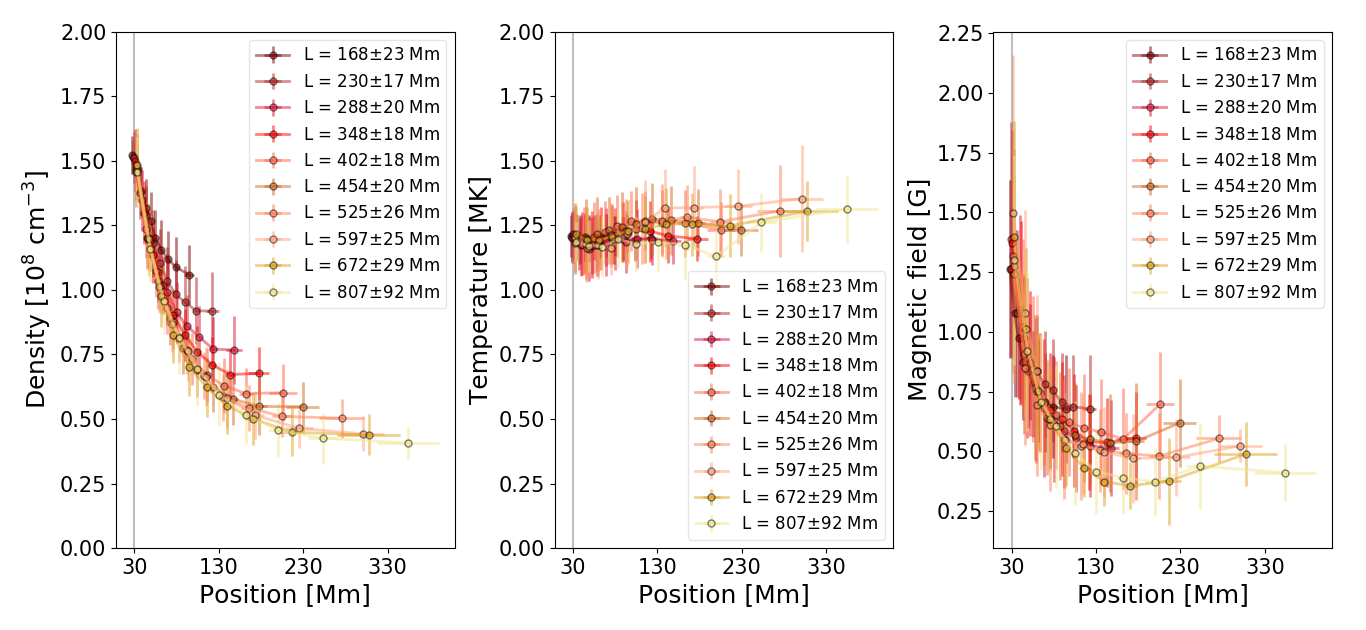}
\caption{Density (left panel), temperature (central panel), and magnetic field (right panel) profiles for each reconstructed typical loop leg. The error bars correspond to the standard deviation of the mean values computed at 10 equidistant positions along the loops (see text). {The grey vertical line is set at $30\,\rm{Mm}$ to indicate the approximate location of the first data points.}}
\label{NTB}
\end{center}
\end{figure*}

Figure~\ref{NTB} shows the profiles of median temperature, electron density, and magnetic field of the typical loop legs just described. {The first data points of all profiles lie around $\approx 28-33\,\rm{Mm}$ depending on the loop leg length. A grey vertical line set in $30\,\rm{Mm}$ is added to indicate the approximate starting point.} We found a similar behavior for all the typical loops. Densities (see Figure~\ref{NTB}, left panel) present a decreasing exponential-like behavior as expected. While shorter loops have higher median densities at all positions along the legs, the longer loops show very similar density profiles. This is a consequence of the non-semicircularity of the longer tomographic loops. As we choose them to be within the tomographic limits, some of the largest loops tend to be flat at the top and thus, loop maximum heights are similar, making them present almost constant density profiles at their tops. As we will see, this is a key point in the comparison of the typical loops with the hydrodynamic model in Section~\ref{EBTELDEMT}.

Magnetic fields (see Figure~\ref{NTB}, right panel) also present a decreasing profile and typical values expected for the quiescent corona. It can be seen in Figure~\ref{NTB} that the maximum value of the magnetic field does not exceed the $2\,\rm{G}$.

The small variation observed in the temperature profiles (see Figure~\ref{NTB}, central panel) are well within the characteristic size of the error bars ($\approx0.25\,\rm{MK}$), which are defined by the median of the temperature width provided by the tomographic procedure (i. e., the width of the LDEM distribution, see Section~\ref{DEMT}). This indicates that although the profiles of the median temperature of the tomographic loops is approximately constant, individual loops have relatively wide thermal distributions.

In a previous work we studied CR 2082 searching for scaling laws between different properties of loops reconstructed with the tomography technique \citep{maccormack_2020}. 
{As we mentioned in Section~\ref{sec1}, in that work we found scaling laws between loop-average plasma properties and loop lengths obtained from the direct analysis of the full loop set.  Here we use the different approach of constructing typical loops as described previously in this section. Our first step is to check that the new data arrangement in the form of typical loops approximately reproduces the scaling laws found in \citet{maccormack_2020}.}

{It is also worth to note that despite the fact that we used the tomographic technique on the same CR in both studies,} the set of loops is not the same, because in our previous work we imposed different conditions on the loop selection. In addition to those used in this work (good distribution of data in each analyzed leg), in \citet{maccormack_2020} we required that density and temperature profiles present good exponential and linear fits respectively. These two latest criteria were discarded in this work in order to expand the data set. We refer the reader to \citet{maccormack_2020} for detailed information on the imposed requirements.

As a brief summary, in \citet{maccormack_2020} we computed, for each tomographic loop, the median values of their thermal and magnetic properties averaged along the loop length (henceforth, loop-average values). We found that the loop-average density decreases with length as $N_m \approx L^{-0.35}$, and the loop-average magnetic field has a relation with the loop length of the form $B_m \approx L^{-r}$, where $r$ must be $0.3$ in order to agree with the radiative flux approximation, $\phi_r \approx N_m L B^{-1}$, derived from Equation~\ref{phir}. The value of $r$ that we found was in the range $[0.15,0.55]$. We did not find any direct relation between the loop-average temperature and the loop length. 

{In order to check if the typical loops constructed here follow the same scaling laws found in \citet{maccormack_2020},} we compute their loop-average density, temperature, and magnetic field. {Here, we define ``loop-average'' as the median of the discrete values that the thermal properties take along the loop.} We consider the standard deviations as the corresponding errors for each data point. 

The upper panel of Figure~\ref{NTBvsL} shows the loop-average density of the typical loops as a function of length. We see that it decreases with length as expected, because longer loops tend to reach larger heights; then, have lower average densities than loops closer to the coronal base, as it can be clearly seen in the leftmost panel of Figure~\ref{NTB}. Also, longer loops tend to have larger error bars due to the wider variation of density. We perform a least square fit of the data finding an exponent of $-0.36$ in agreement with our previous results.  

The relation between loop-average temperature and loop length for the typical loops is shown in the middle panel of Figure~\ref{NTBvsL}. We obtain an approximately null exponent from the least square fit, given by the behavior of the typical loops which are, in average, isothermal, as clearly seen in the middle panel of Figure~\ref{NTB}. As we mentioned before, all temperatures are within the limits of the sensitivity range of the EUV telescope used.

The lower panel of Figure~\ref{NTBvsL} shows the relation between the loop-average magnetic field and the loop length. We find an exponent of $r\approx0.24$, similar to what we found in our previous work. 

\begin{figure}[ht!]
\begin{center}
\includegraphics[width=0.42\textwidth]{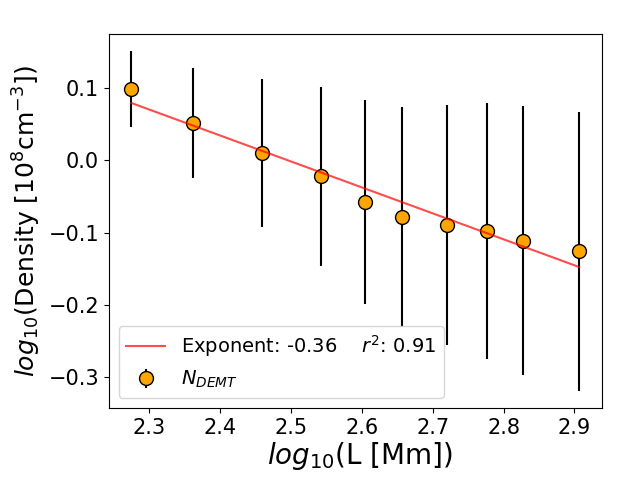}
\includegraphics[width=0.42\textwidth]{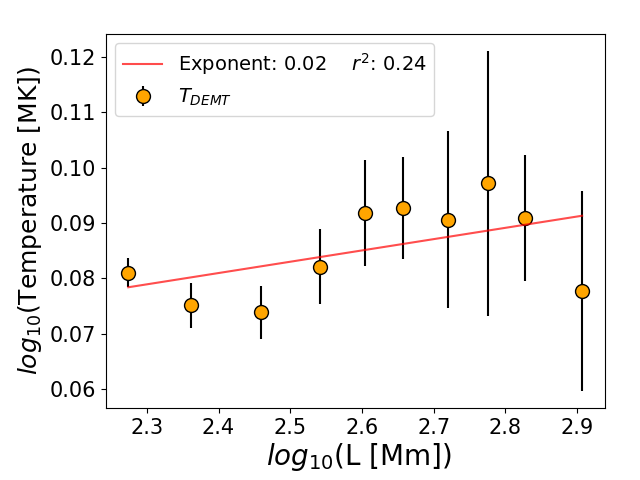}
\includegraphics[width=0.42\textwidth]{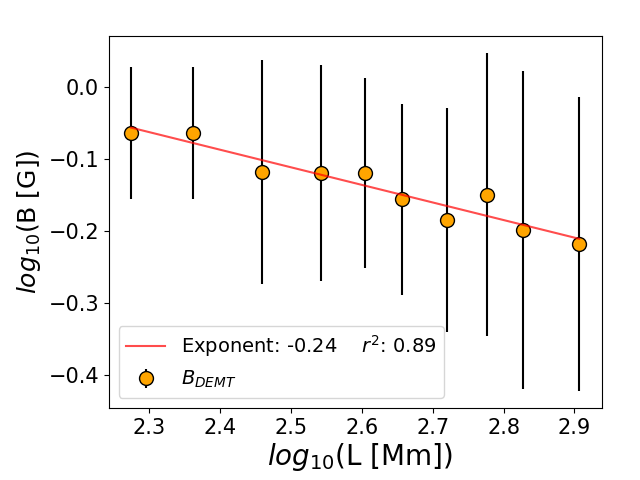}
\caption{Scatter plots of loop-average densities (upper panel), temperatures (middle panel), and magnetic fields (lower panel) vs. loop length for the studied typical loops. Orange dots represent loop-average properties starting from a height of $\approx17\,\rm{Mm}$. Error bars are the standard deviation of the parameters of each typical loop. Continuous lines correspond to a linear least-square fit of the data. The fitting parameters and their corresponding $r^2$ coefficients are provided in the insets.}
\label{NTBvsL}
\end{center}
\end{figure}

Using the energy balance model described at the end of Section~\ref{DEMT}, we compute the energy input flux at the coronal base for each typical loop. In Figure~\ref{HvsL} we plot these fluxes as a function of loop length. We find a relation $\phi_h \approx L^{0.55}$. The exponent is at the lower end of the range obtained in \citet{maccormack_2020}, where we found exponents within the interval $[0.55,0.84]$, depending on the solar latitude location of the studied loops.

\begin{figure}[ht!]
\begin{center}
\includegraphics[width=0.42\textwidth]{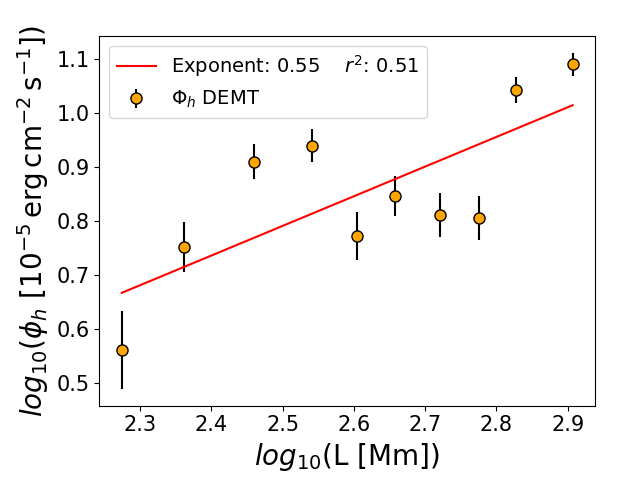}
\caption{Same as Figure~\ref{NTBvsL} for the loop-integrated energy input flux $\phi_h$.}
\label{HvsL}
\end{center}
\end{figure}

{Since at heights well above the transition region the temperature gradients are very small, conductive fluxes are one order of magnitude smaller than typical radiative fluxes. Then, most of the energy input flux in the coronal part is determined by the radiative flux. This has been discussed in \citet{maccormack_2020}. Because the radiative flux depends strongly on the density, the closer to the TR, the larger the emission and its contribution to the heating flux. This also implies that most of the emission occurs at temperatures below $0.75\,\rm{MK}$ \citep[see e.g.]{cargill_2012}. The later temperature ranges, and heights at which they occur (below 17 Mm), are not included in the computation of the tomographic technique, since the signal to noise relation is not good at low heights and the technique does not reconstruct the plasma properties in a completely reliable way.}

\section{Typical loop modeling}
\label{EBTELDEMT}

To study the thermal state of the typical loops obtained in Section~\ref{RL}, first we use EBTEL to model the loop-average densities and temperatures that a typical loop should have if the energy input was the heating flux computed with the simple energy balance model described in Section~\ref{DEMT}. This heating flux is transformed into the volumetric heating rate needed as input for EBTEL by assuming that it is uniform along the loop and dividing it by the loop length. With this heating rate we run EBTEL until the density and temperature become stationary.

For a quantitative comparison, we compute the ratio between the loop-average densities and temperatures of each typical loop (TL) and the values obtained with EBTEL. {Results of the comparison are shown in Table}~\ref{EBDCOM}. We find that the mean value of the density ratio for all the typical loops is $<N_{TL}/N_{EBTEL}>=0.67$, with a standard deviation of 0.13, while the mean of the temperature ratio is $<T_{TL}/T_{EBTEL}>=0.53$, with a standard deviation of 0.11. 

\begin{table}[ht]
\begin{center}
\begin{tabular}{ c | c | c }
 $L$ & $N_{TL}/N_{EBTEL}$ & $T_{TL}/T_{EBTEL}$ \\ 
 \hline
 \hline
 188 & 0.88 & 0.78 \\ 
 230 & 0.51 & 0.66 \\ 
 288 & 0.52 & 0.53 \\ 
 348 & 0.53 & 0.51 \\ 
 402 & 0.74 & 0.56 \\ 
 454 & 0.68 & 0.51 \\ 
 525 & 0.79 & 0.51 \\ 
 597 & 0.87 & 0.50 \\ 
 672 & 0.60 & 0.41 \\ 
 807 & 0.62 & 0.36 \\ 
\end{tabular}
\end{center}
\caption{Ratio computed between loop-averaged density $N_{TL}$ (temperature $T_{TL}$) for each TL and loop-average EBTEL density $N_{EBTEL}$ (temperature $T_{EBTEL}$) for each TL length.}
\label{EBDCOM}
\end{table}

The results presented in Table~\ref{EBDCOM} show that the simple energy balance model and the EBTEL code do not reproduce the same thermal conditions for loops assumed to be in equilibrium. This shows that the assumed equilibrium does not reflect the actual conditions of the quiescent corona and that some kind of dynamics might be present in the loops that remains hidden by the averaging of the tomographic technique.

To understand these results we study two alternative scenarios in a similar way to what we did in \citet{nuevo_2020} for active region loops. In the first scenario, we assume again loops in equilibrium and heated with an energy input which is constant in time and uniform throughout the loop, but this time we provide EBTEL with different volumetric heating rates until we find the one that best reproduces the loop-average temperature of the typical loops. We then check if the resulting densities correspond to the loop-average densities of the TLs. 

In the second scenario, we consider the average thermal properties of the loops evolving dynamically due to a variable heating. In particular, we model loops heated by nanoflare-like short duration impulsive events. The thermal properties provided by the tomographic technique would correspond, in this case, to the averages of densities and temperatures that evolve in time. As before, we try different energy levels until we find the best match with the TL temperatures and then we analyze how well the mean densities reproduce the TL values.

We model the impulsive events as triangular functions defined by a heating rate peak ($E_{imp}$), and a duration ($\tau$). Considering impulsive events produced by magnetic stress fed by photospheric motions, as proposed by \citet{parker_1988}, we choose $\tau\le 300$ s, in correspondence with a characteristic time of the photospheric granulation, i.~e., the typical duration of a photospheric granule. In general, during the impulsive phase of a nanoflare the plasma reaches temperatures which are higher than the sensitivity window of the EUVI instrument used here. Then, for the average computations we consider the cooling phase of the nanoflare, when the temperature lies within the limits of the instrumental sensitivity range $[0.5,2.5]\,\rm{MK}$. We average the temperature and the density along the time interval ($\tau_e$) during which the temperature of the plasma remains within that range. 

It is worth noting that the averaged physical conditions of the evolving plasma could correspond to a monolithic loop or a loop formed by sub-resolution elemental magnetic strands, as proposed by \citet{parker_1988}. The modeling and averaging performed here is just to test if impulsive heating better reproduces the mean thermal properties of the constructed TLs. The statistical nature of this analysis does not allow us, at this point, to discriminate one scenario (monolithic loops) from the other (multi-stranded loops). 

In order to correctly represent the evolution of a nanoflare, we first need to impose realistic initial conditions. Since EBTEL provides, by default, initial conditions which might not be typical of a coronal state, we precede the test nanoflare by two other identical events that provide a consistently initial state.

We apply the constant and impulsive heating scenarios just described to model the TLs presented in Section~\ref{RL}. For each TL we use as EBTEL input the loop length and we adjust the heating to reproduce the loop-average temperature. We then compare the mean coronal density obtained with the loop-average density of the TL. 

In Table~\ref{input-par} we present the list of parameters used to model the loops using constant and impulsive heating. The first column shows the typical loop lengths ($L$), the second column corresponds to the heating rate applied in the constant heating case ($E_{const}$), and the third to fifth columns show, respectively, the peak heating rate of the nanoflare ($E_{imp}$), its duration ($\tau$) and the time during the nanoflare cooling phase ($\tau_e$), along which the temperature of the plasma is within the sensitivity window of the EUVI instrument used.


\begin{table}[ht]
\begin{center}
\begin{tabular}{ c | c | c c c}
 $L$ & $E_{const} \times10^{-6}$ & $E_{imp}$ & $\tau$ &$\tau_e$ \\ 
 $[\rm{Mm}]$ & $[\rm{erg}\,\rm{cm}^{-3}\,\rm{s}^{-1}]$ & $[\rm{erg}\,\rm{cm}^{-3}\,\rm{s}^{-1}]$ & $[s]$ & $[s]$ \\ 
 \hline
 \hline
 188 & 7.5 & 0.04 & 200 & 7500  \\ 
 230 & 5.7 & 0.03 & 50  & 7500  \\ 
 288 & 2.9 & 0.07 & 200 & 14100  \\ 
 348 & 2.2 & 0.06 & 300 & 16000  \\ 
 402 & 1.8 & 0.08 & 50 & 18000  \\ 
 454 & 1.4 & 0.10 & 50 & 20600  \\ 
 525 & 1.2 & 0.10 & 300 & 22500  \\ 
 597 & 1.0 & 0.14 & 200 & 25000  \\ 
 672 & 0.7 & 0.15 & 200 & 28500  \\ 
 807 & 0.4 & 0.17 & 300 & 35000  \\ 
\end{tabular}
\end{center}
\caption{EBTEL input parameters used to model each typical loop with both heating mechanisms. We show the loop length, $L$, the heating rate for constant ($E_{const}$) and impulsive ($E_{cont}$) energy injections, duration of the impulsive events ($\tau$) and duration of the interval along which the thermal properties are averaged $\tau_e$.}
\label{input-par}
\end{table}

In Table~\ref{comp} we compare the loop-average densities computed with EBTEL for constant and impulsive heating with those corresponding to the TL. For the comparison we use the ratio between the loop-average densities of the TLs divided by the EBTEL corresponding values.  

The results clearly indicate that impulsive heating models reproduce better the reconstructed typical loop densities, in particular, for the shorter loops up to approximately 600 Mm. We can see that longer loops tend to need longer evolution times ($\tau_e$) and higher heating intensities, to be close to the TL values. It can also be noticed that for the longer loops the EBTEL modeled density tends to depart from the TL values. The possible reasons for this are discussed in Section~\ref{conc}.   

{It is noteworthy that for TLs with $L$\,=\,230\,\rm{Mm}, 402\,\rm{Mm}\,\rm{and} 454\,\rm{Mm}, the $\tau$ times are markedly shorter than for other lengths. These simply happen to be the impulsive event times for which the modeling provides a better approximation to the TL loop-average plasma properties. Clearly, this is due to the peculiarities of the density and temperature distributions along these loops. An analysis of these peculiarities would require a more thorough 1D modeling of the loops (see Section~\ref{conc}).}

\begin{table}[ht]
\begin{center}
\begin{tabular}{ c | c | c }
 $L$ & $\langle N_{TL}\rangle/\langle N_{EBTEL}\rangle_{const}$ & $\langle N_{TL}\rangle/\langle N_{EBTEL}\rangle_{imp}$ \\ 
 \hline
 \hline
 188 & 1.7 & 1.0 \\ 
 230 & 1.6 & 1.0 \\ 
 288 & 3.1 & 1.0 \\ 
 348 & 3.8 & 1.2 \\ 
 402 & 4.5 & 1.1 \\ 
 454 & 5.5 & 1.1 \\ 
 525 & 7.1 & 1.3 \\ 
 597 & 8.3 & 1.3 \\ 
 672 & 14.4 & 1.7 \\ 
 807 & 23.0 & 2.9 \\ 
 \end{tabular}
\end{center}
\caption{Ratio between TL densities ($\langle N_{TL}\rangle$) and those obtained with the EBTEL model for constant ($\langle N_{EBTEL}\rangle_{const}$) and impulsive ($\langle N_{EBTEL}\rangle_{imp}$) heatings.}
\label{comp}
\end{table}

For a graphic representation of the above results, in Figure~\ref{EBTEL2082} we show the loop-average temperature (upper panel) and density (bottom panel) computed with EBTEL for the two alternative heating scenarios: constant (stars) and impulsive (circles), and those corresponding to the TLs (squares). The yellow areas and the orange bars are used as error bar estimations. The yellow areas correspond to the standard deviation of the TL properties, and the orange bars correspond to the standard deviation of the values computed with EBTEL along the time evolution interval, $\tau_e$. The figure clearly shows how the impulsive heating better reproduces the thermal conditions of the reconstructed TLs, at least for loops up to approximately 600 Mm. In Section~\ref{conc} we discuss the implications of our results and we propose future directions to explore.

\begin{figure*}[ht!]
\begin{center}
\includegraphics[width=0.98\textwidth]{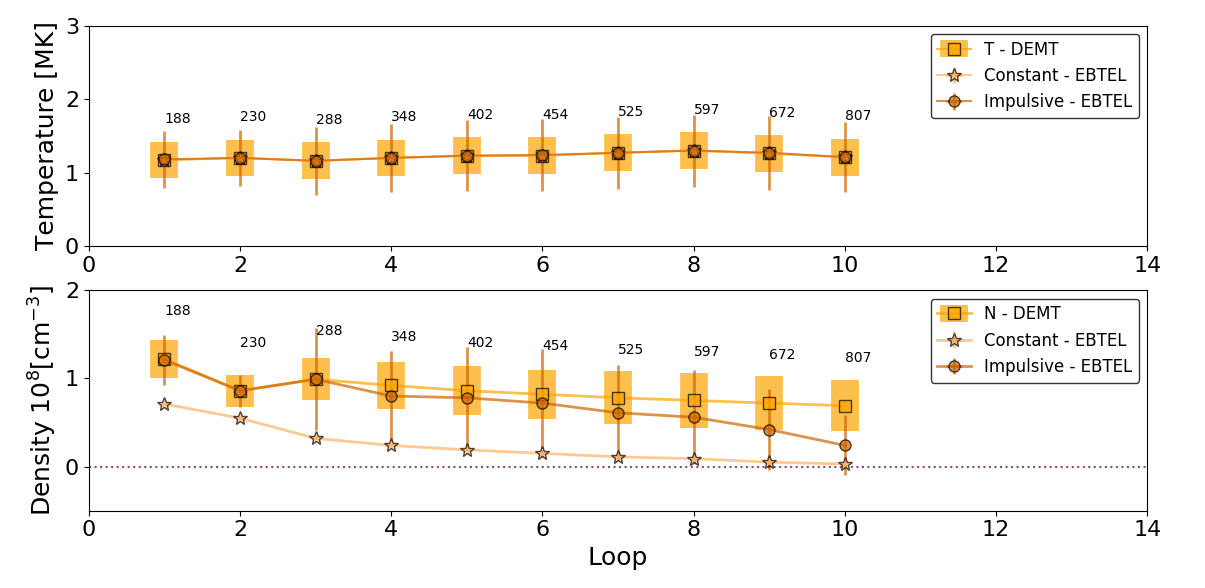}
\caption{Comparison of loop-average temperature (upper panel) and density (lower panel) of the TLs and those obtained with the EBTEL model for the two analyzed heating scenarios: constant (stars) and impulsive (circles). Squares correspond to the loop-average parameters of the TLs and the yellow areas represent their corresponding standard deviations. Orange bars represent the standard deviation of the EBTEL parameters averaged along the time during which the plasma temperature is within the EUVI instrument thermal response range.}
\label{EBTEL2082}
\end{center}
\end{figure*}

\section{Discussion and conclusions}
\label{conc}

In this article we have explored the thermal state of magnetic loops in the quiescent corona, by comparing loops reconstructed with a DEM tomographic method \citep[DEMT, ][]{vasquez_2015} with the Enthalpy Based Thermal Evolution of Loops model \citep[EBTEL, ][]{klimchuk_2008,cargill_2012}. From a full set of approximately 55000 loops reconstructed for Carrington rotation (CR) 2082, we obtain typical loops of ten different lengths between 188 and 807 Mm, following an averaging procedure similar to what is usually done in superposed epoch analysis works. We use EBTEL to model the TLs and show that their mean thermal conditions are not consistent with a state of static or quasi-static equilibrium. We find instead that a heating injection consisting of impulsive nanoflare-like events is more suitable to explain the mean thermal properties of the loops. 

We find that for the longer TLs, above 600 Mm, the densities modeled with impulsive heating tend to progressively depart from the DEMT values. We check the shape of the mean geometric distribution of points along the longer TLs and we find that they progressively depart from the semicircular loop shape assumed by EBTEL, as well as other hydrodynamic models. In particular, the loops tend to take a flatter shape at their tops. This is consistent with the tendency of the longer loops to have constant densities at their tops, as clearly seen in Figure~\ref{NTB}. We think that the shape of the long loops is produced by a selection effect due to our limitation to consider only loops contained within the tomographic limits. Loops of the same lengths with more semicircular shapes are left out of the selection because they exceed tomographic heights. 

{Regarding the PFSS model, it is noteworthy that the magnetic field values provided by the extrapolation at the apex of the longer loops may result in plasma $\beta$ values close to or above 1. Although the PFSS model reproduces successfully a great part of the lower corona, it is not clear yet how well it does for the coronal field at larger heights \citep[see e.g., ][]{riley_2006}. This could imply an important limitation for the loop modeling, because the parallel-to-field hydrodynamic approximation could fail at those plasma $\beta$ values.}

It is perhaps worth to emphasize that our conclusion that impulsively heated loops are more consistent with the thermal conditions of the quiescent corona is not at all definitive, as it is limited by the statistical nature of the present study. DEMT reconstructed loops are, after all, averaged versions of actual loops evolving in timescales which are much shorter than a full solar rotation period. However, although our present analysis cannot reveal the details of the real loop evolutions at their relevant timescales, the results presented here strongly indicate that the quiescent coronal plasma is remarkably overdense with respect to the equilibrium solutions sometimes assumed. In the case of active regions, the overdensity of the coronal plasma has been known for quite some time \citep[see e.g., ][]{klimchuk_2015}, as we recently confirmed with our own analysis in \citet{nuevo_2020}. Similarly to our findings using observed active region loops, here we find a dynamical mechanism that explains the tomographically reconstructed densities and temperatures.

{In a very recent article, published after the present work was submitted, \citet{cargill_2022} presented and studied a new version of the EBTEL code that considers the effect of cross-section variations along the loops. One of their main results is that the mean coronal density of the loops increase when cross-section expansion is included in the computation. The inclusion of this effect in our modeling would clearly improve the  comparison with the typical loop densities for the longer loops. We will address this possibility in a future work.}

Finally, the EBTEL modeling of the analyzed loops is, above all, useful in guiding us on the possible mechanisms, input parameters, and boundary and initial conditions that could be used in future more sophisticated studies. {Since EBTEL computations are based on mean coronal values, the model does not resolve the energy equation in an exact way, so errors can be carried on along the computations. It is also necessary to remark that the EBTEL model has been tested against a 1D hydrodynamic model only for loops as long as $\approx 200\,\rm{Mm}$ \citep[see e.g., ][]{cargill_2012,cargill_2022}. Longer loops may require a more complex stratification than the one used by EBTEL. Although we plan to thoroughly explore this in the future, all these caveats need to be considered regarding the present conclusions.} The main advantage of using EBTEL is its non-demanding computational time that makes it possible to perform many different runs in reasonable times. Future work will focus on using 1D hydrodynamic models to reproduce the full typical loop leg profiles constructed here. 

\section{Acknowledgments}
The authors sincerely thank the two reviewers for enriching comments and suggestions that greatly contributed to improve the article. The authors acknowledge financial support from the Argentine grants PICT 0221 (ANPCyT) and UBACyT 20020130100321. MLF, CHM and AMV are members of the Carrera del Investigador Cient\'ifico of the Consejo Nacional de Investigaciones Cient\'ificas y T\'ecnicas (CONICET). CMC and DGL are PhD fellows of CONICET.

\appendix

\section{}
\label{ADEMT}

The DEMT technique was initially developed by \citet{frazin_2009} and improved in later works \citep[see ][ for a review]{vasquez_2015}. It is a global technique that reconstructs the 3D distribution of the thermal properties of the quiescent corona from a series of EUV images in different bands covering a full solar rotation. 

The DEMT divides the inner corona, between the heights of $17-174\,\rm{Mm}$, in a spherical grid. In this work we use grid voxels of $\approx 7\,\rm{Mm}$ in the radial direction and $2^{\circ}$ in latitude and longitude. With this dimensions, the technique requires a cadence of one image every 6 hours, resulting in a total of 110 images covering a complete solar rotation. 

The firs step is to solve an inversion problem that from a series of EUV images produces a 3D distribution of the filter band emissivity (FBE) for each EUV band used. This FBE is the integral over wavelength of the EUV spectral emissivity and the telescope's passband function for each band. 

The second step is to use the FBE values for the determination of the local differential emission measure (LDEM) in each tomographic voxel. When three coronal bands are used to compute the FBE, LDEM is commonly modeled with a Gaussian function \citep{nuevo_2015} that depends on three free parameters: centroid, width and total area. These three parameters are related with the thermal properties of the plasma present in each tomographic voxel. The technique selects the three parameters that better globally reproduce the FBE values mentioned before. 

Finally, by taking the moments of the LDEM in each voxel $i$, we obtain the electronic density $N_{e,i}$, the mean temperature $T_{e,i}$ and the temperature distribution width, $WT_{i}$, which gives an indication of how multithermal the plasma in the voxel is. In this way we obtain a 3D distribution of the thermal properties of the coronal plasma within the limits of the tomographic grid.

For more details, we refer the reader to \citet{frazin_2009} and \citet{vasquez_2015}.

The obtained DEMT results are then combined with a potential-field source-surface (PFSS) magnetic model \citep{huang_2012} in the following way. As a first step, we extrapolate the magnetic field from a photospheric synoptic magnetogram corresponding to the CR of interest. Here, we use a synoptic magnetogram obtained with the Michelson Doppler Imager (MDI), on board the Solar and Heliospheric Observatory, \citep[SOHO,]{scherrer_1995}. We set the source surface of the model at a height of $2.5\,R_{\odot}$. 

Once the model is obtained, coronal magnetic field lines are integrated from starting points located at the center of each tomographic voxel. Each integrated field line then crosses several voxels of the grid. Since the spatial resolution of the PFSS model is higher than the tomographic one, of all the points of a magnetic field line that lie within an individual voxel, we select the median one and we assign to it the LDEM moments $N^2_{e,i}$ and $T_{e,i}$ of the voxel. In this way, we can now track thermodynamic properties along magnetic lines and identify each field line with a coronal loop.

\bibliographystyle{model5-names}
\biboptions{authoryear}
\bibliography{refs}

\end{document}